\documentclass[]{interact}

\usepackage{epstopdf}
\usepackage[caption=false]{subfig}
\usepackage[numbers,sort&compress]{natbib}
\usepackage{hyperref}
\usepackage{url}
\usepackage{xcolor}

\bibpunct[, ]{[}{]}{,}{n}{,}{,}

\newcommand{\m}{\mathbf}

\begin{document}

\articletype{Regular article}

\title{Observation by SANS and PNR of pure N\'eel-type domain wall profiles and skyrmion suppression below room temperature in magnetic [Pt/CoFeB/Ru]$_{10}$ multilayers}

\author{
\name{Victor Ukleev\textsuperscript{a,b}\thanks{CONTACT Victor Ukleev. Email: victor.ukleev@helmholtz-berlin.de}, Fernando Ajejas\textsuperscript{c}\thanks{CURRENT ADDRESS Fernando Ajejas: IMDEA Nanociencia, C/ Faraday 9, Campus de Cantoblanco}, Anton Devishvili\textsuperscript{d}, Alexei Vorobiev\textsuperscript{d,e}, Nina-Juliane Steinke\textsuperscript{d}, Robert Cubitt\textsuperscript{d}, Chen Luo\textsuperscript{b}, Radu-Marius Abrudan\textsuperscript{b}, Florin Radu\textsuperscript{b}, Vincent Cros\textsuperscript{c}, Nicolas Reyren\textsuperscript{c}, Jonathan S. White\textsuperscript{a} \thanks{CONTACT Jonathan S. White. Email: jonathan.white@psi.ch}}
\affil{\textsuperscript{a}Laboratory for Neutron Scattering and Imaging (LNS), Paul Scherrer Institute (PSI), CH-5232 Villigen, Switzerland; \\ 
\textsuperscript{b}Helmholtz-Zentrum Berlin f\"ur Materialien und Energie, D-12489 Berlin, Germany; \\ 
\textsuperscript{c}Unit\'e Mixte de Physique, CNRS, Thales, Universit\'e Paris-Saclay, 91767 Palaiseau, France;\\
\textsuperscript{d}Institut Laue-Langevin, 71 Avenue des Martyrs, 38042 Grenoble, France; \\
\textsuperscript{e}Department of Physics, Uppsala University, Uppsala SE-751 20, Sweden;}}

\maketitle

\begin{abstract}
We report investigations of the magnetic textures in periodic multilayers [Pt(1 nm)/(CoFeB(0.8 nm)/Ru(1.4 nm)]$_{10}$ using polarised neutron reflectometry (PNR) and small-angle neutron scattering (SANS). The multilayers are known to host skyrmions stabilized by Dzyaloshinskii-Moriya interactions induced by broken inversion symmetry and spin-orbit coupling at the asymmetric interfaces. From depth-dependent PNR measurements, we observe well-defined structural features, and obtain the layer-resolved magnetization profiles. The in-plane magnetization of the CoFeB layers calculated from fitting of the PNR profiles is found to be in excellent agreement with magnetometry data. Using SANS as a bulk probe of the entire multilayer, we observe long-period magnetic stripe domains and skyrmion ensembles with full orientational disorder at room temperature. No sign of skyrmions is found below 250\,K, which we suggest is due to an increase of an effective magnetic anisotropy in the CoFeB layer on cooling that suppresses skyrmion stability. Using polarised SANS at room temperature, we prove the existence of pure N\'eel-type windings in both stripe domain and skyrmion regimes. No Bloch-type winding admixture, i.e. an indication for hybrid windings, is detected within the measurement sensitivity, in good agreement with expectations according to our micromagnetic modelling of the multilayers. Our findings using neutron techniques offer valuable microscopic insights into the rich magnetic behavior of skyrmion-hosting multilayers, which are essential for the advancement of future skyrmion-based spintronic devices.
\end{abstract}


\section{Introduction}

Skyrmions are spin whirling particle-like textures that attract significant attention due to their topological nature and outstanding potential for use as information bits in spintronic devices \cite{bogdanov1994thermodynamically,sampaio2013nucleation,everschor2018perspective,song2020skyrmion}. Magnetic multilayers (MMLs) composed of alternating ferromagnetic / heavy-metal layers are one of the most promising classes of skyrmion-hosting systems due to their high tunability via layer engineering. The combination of broken inversion symmetry and spin-orbit coupling at the asymmetric MML interfaces induces the Dzyaloshinskii-Moriya interaction (DMI), resulting in the stabilization of small, mobile N\'eel-type skyrmions at room temperature \cite{moreau2016additive, woo2016observation, fert2017magnetic, jiang2017skyrmions, soumyanarayanan2017tunable}. Recently, it was discovered that despite the thinness ($\sim$~1~nm or less) of the magnetic layers, the magnetic order in MMLs is also affected by dipolar interactions, which favours Bloch-type domain walls and skyrmions in MMLs with increasing number of multilayer periods \cite{legrand2018hybrid}. The competition between DMI and dipolar energies results in the formation of spin textures with mixed N\'eel-Bloch i.e. hybrid windings, with this phenomenon already observed in previous resonant x-ray scattering studies performed on relatively thick magnetic films \cite{legrand2018hybrid,zhang2018reciprocal,li2019anatomy}. While undoubtedly a powerful probe of the microscopic winding-type of mesoscale spin textures, resonant x-ray scattering is inherently limited by the low penetration depth of soft x-rays of $\sim$10-100\,nm, which provides the view of the surface region but not the entire MML \cite{legrand2018hybrid,zhang2018reciprocal,ukleev2022chiral,grelier2022three,burgos2023probing}. In contrast, neutron scattering is a bulk-sensitive probe that generally allows the investigation of cm-thick specimens installed in complex sample environments \cite{muhlbauer2019magnetic}. Therefore, neutron techniques provide access to the magnetic order throughout entire MML samples, providing a global view of the microscopic magnetism that complements the probes most sensitive to the surface region. In addition, neutron scattering experiments with polarisation analysis allow a detection of N\'eel- and Bloch-type domain walls within the entire MML system, a property that is largely inaccessible in relatively thick samples for other experimental techniques \cite{desautels2019realization,tang2023skyrmion,liyanage2023three}.

In the present work we investigate the [Pt(1 nm)/(CoFeB(0.8 nm)/Ru(1.4 nm)]$_{10}$ MML using a powerful combination of polarised neutron reflectometry (PNR), and both unpolarised and polarised small-angle neutron scattering (SANS). As our main results, we show that the entire MML with 10 repetitions is unambiguously characterized by purely N\'eel-type domain walls in both zero-field stripe domain (cycloidal) and field-induced skyrmion phases. Further, we show that the skyrmion phase does not extend towards low temperatures below 250~K, presumably due to the development of a destabilising effective anisotropy on cooling below room temperature. In combination with micromagnetic modelling, our experimental findings help to refine the magnetic parameters for the MMLs and provide valuable microscopic feedback for both theoretical studies and rational MML design.

\section{Materials and methods}

Ta(5 nm)/Pt(7 nm)/[Pt(1 nm)/Co$_{40}$Fe$_{40}$B$_{20}$(0.8 nm)/Ru(1.4 nm)]$_{10}$/Pt(3 nm) multilayers were synthesized by room-temperature magnetron sputtering on thermally oxidized Si substrates 2" in diameter. The Si substrate is (0 0 1)-oriented, but due to the oxidized layer the surface of the substrate is amorphous. However, due to the sputtering growth process, the buffer Pt layer grows preferentially along the (1 1 1) direction and forms grains of a few nm in diameter \cite{mcvitie2018transmission}. The Ru layer is used for stabilizing the growth quality of the MML and to dissymmetrize the Pt|CoFeB interface with relatively strong DMI. The Pt layer strongly suppresses the Ruderman-Kittel-Kasuya-Yosida (RKKY) electronic interlayer coupling induced by the Ru (ferromagnetic for 1.4\,nm thick Ru thickness). More details on tuning the interactions via layers engineering are given in Refs. \cite{legrand2020room,legrand2022spatial,ajejas2023densely}.

\begin{figure*}
\includegraphics[width=1\linewidth]{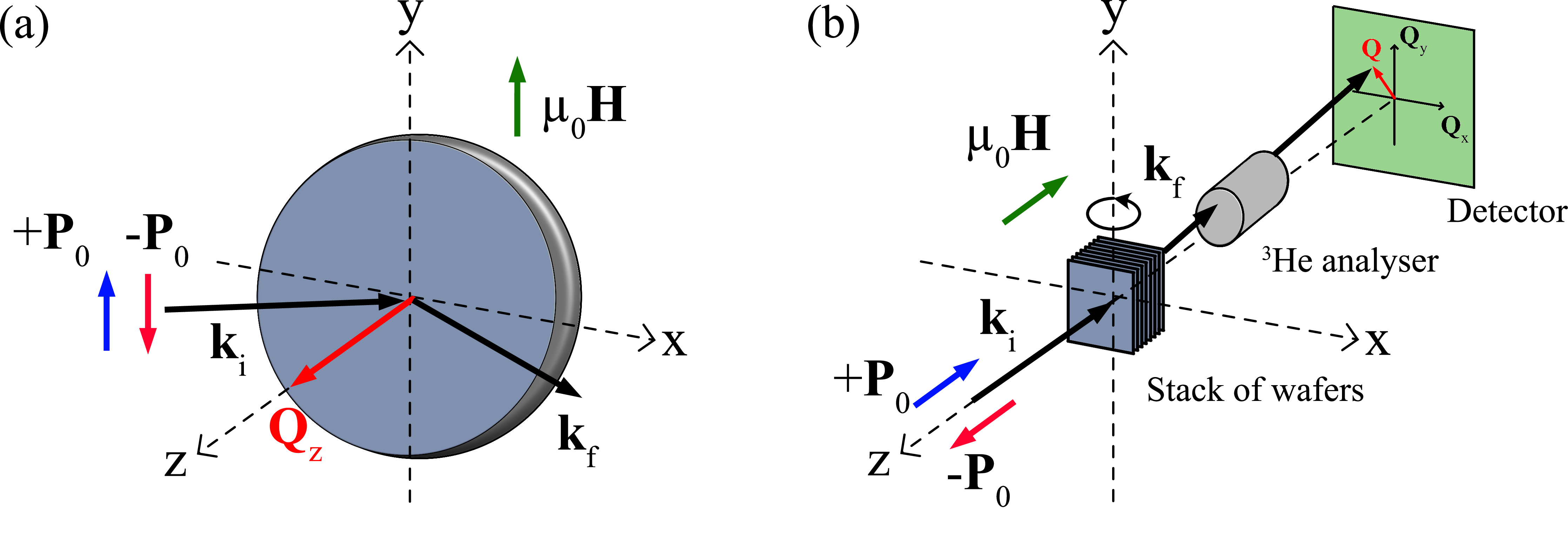}
        \caption{Schematics of (a) polarised neutron reflectometry (PNR), and (b) small-angle neutron scattering (SANS) experimental geometries. In the case of unpolarised SANS experiments an unpolarised neutron beam was used and the $^3$He analyser was not present.}
        \label{fig1}
\end{figure*}

For the PNR experiment, a single MML was used. The PNR measurements were carried out at the the Institut Laue-Langevin (ILL), Grenoble, France using the SuperADAM reflectometer \cite{vorobiev2015recent} with a horizontal scattering plane and a neutron wavelength of 5.21\,\AA. A schematic illustration of the PNR experimental geometry is shown in Fig. \ref{fig1}(a). Using a supermirror polariser, the incident neutron beam was polarised vertically ($P_0=99.8\%$), and perpendicular to the scattering plane. No polarization analysis was employed. The experiment was carried out at room temperature. Vertical magnetic fields $\mu_0 H_y$ in range from 1\,mT to 700\,mT were applied parallel (or anti-parallel) to the incoming neutron polarization vector, $\mathbf{P}_0$ using an electromagnet.

Unpolarised SANS measurements were performed at the ILL using the D33 instrument \cite{dewhurst2016small} with a neutron wavelength of 9\,\AA. To increase the scatting volume in the SANS experiment, identical MML samples were cut into $15\times15$\,mm$^2$ square pieces, and stacked into a bespoke sample holder \cite{kanazawa2016direct,kanazawa2020direct}. 29 pieces were stacked in total, thus yielding a total thickness of $\sim$210\,nm of CoFeB along the neutron beam path. The stack was mounted in a horizontal-field cryomagnet that controlled the temperature and magnetic field. A schematic illustration of the SANS experiment is shown in Fig. \ref{fig1}(b). For the measurements, the magnetic field was parallel to the incident neutron wavevector $\mathbf{k}_i$. The sample could be rotated around the $y$-axis inside the magnet in order to perform an in-plane field training procedure.

SANS experiments with polarization analysis were also carried out at D33 using 10\,\AA~neutrons. The incident beam was polarised using a supermirror (polarization of the incoming beam $\mathbf{P}_0 || \mathbf{k}_i$, $P_0=96\%$ ) while a $^3$He cell placed between the sample and detector was used for analysing the polarization of the scattered neutrons. A guide magnetic field of 20\,mT was applied in the polarised SANS measurement to maintain the minimal flipping ratio (FR) of 19 throughout the setup. The FR was measured periodically, and corrections due to imperfections in the polarising efficiency of the the entire setup are applied to the scattered SANS intensities in the data analysis. The background scattering for both unpolarised and polarised SANS experiments was measured in the saturated state of the MML at $\mu_0 H_z = 220$\,mT and subsequently subtracted. The typical SANS acquisition time was 45 minutes per pattern with unpolarised beam. In the polarised SANS measurement patterns were acquired for 60 minutes and 180 minutes for non-spin-flip (NSF) and spin-flip (SF) channels, respectively. All SANS data were analysed using the GRASP software package developed at the ILL \cite{dewhurst2023graphical}.

\section{Results and discussion}
\subsection{Magnetic depth profiles}

\begin{figure*}
\includegraphics[width=1\linewidth]{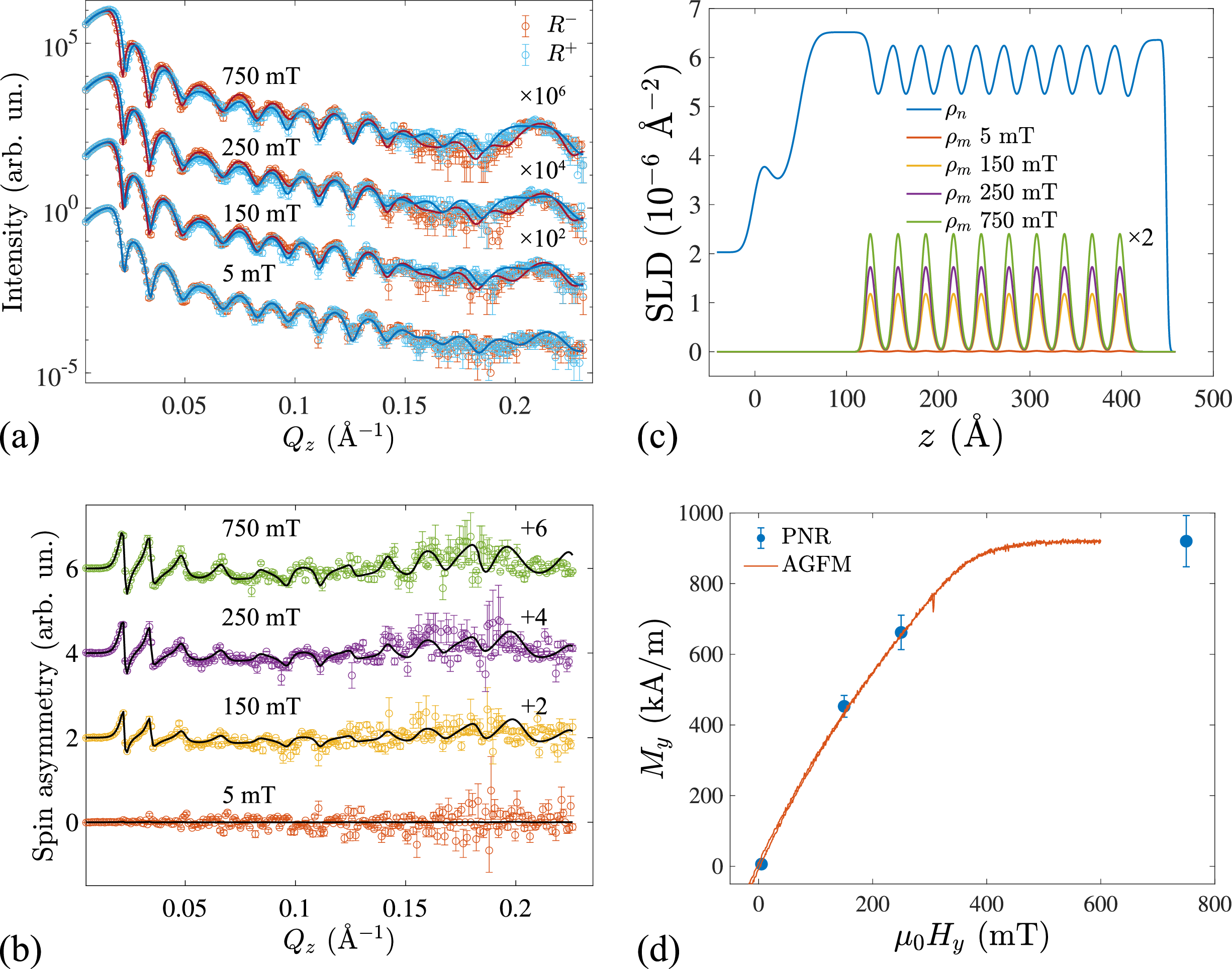}
        \caption{(a) $R^-$ and $R^+$ PNR curves (symbols) measured from Ta(5 nm)/Pt(7 nm)/[Pt(1 nm)/CoFeB(0.8 nm)/Ru(1.4 nm)]$_{10}$/Pt(3 nm) sample at 5, 150, 250, and 750\,mT. Solid lines correspond to the fitted model. Curves measured at different fields are shifted along the $y$-axis for clarity (note the log scale). (b) Measured (symbols) and fitted (solid lines) spin asymmetry ratios for the corresponding magnetic fields.} (c) Reconstructed nuclear ($\rho_n$) and magnetic ($\rho_m$) SLD profiles of the multilayer. $z=0$ corresponds to the interface between the MML and the Si substrate. The magnetic $\rho_m(z)$ profiles are multiplied by a factor of 2 for clarity. (d) Average magnetization ($M_y$) of CoFeB layers as a function of the in-plane magnetic field, as obtained from the fit. Scaled in-plane Alternating Gradient Field Magnetometry (AGFM) data are included for comparison.
        \label{fig2}
\end{figure*}

Using PNR, we investigated the magnetic multilayer sample to obtain layer-resolved magnetization profiles. The specular reflectivity components $R^{+}$ and $R^{-}$, measured with $+\mathbf{P}_0$ or $-\mathbf{P}_0$ incoming polarizations are shown in Fig. \ref{fig2}(a). Each curve represents a superposition between the nuclear and magnetic neutron scattering intensities. Well-pronounced thickness oscillations (Kiessig fringes) and a multilayer Bragg peak at $Q_z=0.21$\,\AA$^{-1}$ are clearly observed indicating the excellent structural quality of the sample \cite{zhu2005modern}.

The magnetic contribution to $R^{+}$ and $R^{-}$ reflectivity curves is determined by the in-plane magnetization component of the sample which is parallel to the incoming neutron polarization vector $\mathbf{P}_0$. The total difference between the $R^{+}$ and $R^{-}$ curves is proportional to the net magnetization of the sample. To observe the difference in $R^+$ and $R^-$ curves related to the magnetic scattering signal more clearly, Fig. \ref{fig2}(b) shows the corresponding spin asymmetry curves. By fitting the PNR curves using GenX$^3$ software \cite{glavic2022genx}, we reconstructed both the nuclear ($\rho_\textrm{n}$) and magnetic ($\rho_\textrm{m}$) scattering length density (SLD) profiles of the MML (Fig. \ref{fig2}(b)). The fitted thicknesses of Pt $d_{\textrm{Pt}}=1.2\pm0.2$\,nm, Ru $d_{\textrm{Ru}}=1.2\pm0.2$\,nm and CoFeB $d_{\textrm{CoFeB}}=0.6\pm0.2$\,nm layers in the periodic stack match the nominal values aimed at in the sputtering process quite well. Peaks and dips in the nuclear SLD profile of the periodic structure correspond to Pt and Ru layers, respectively. The results further reveal a periodic magnetization profile, somewhat smoothed at the interfaces by the conformal roughness. Including a finite proximity-induced moment on Pt \cite{lee2019enhanced,hauser2023spin} and Ru layers does not improve the fit, hence, we assume any induced magnetization lies below our measurement sensitivity. Importantly, the fitted model suggests a similar magnetization of individual CoFeB layers within the stack, thus validating the previously assumed homogeneity of the layer magnetization in the MMLs. The magnetization of the CoFeB layers along the applied field is calculated from the magnetic SLD to be $M_y\textrm{[kA/m]}=3.5\cdot10^9\rho_m\textrm{[\AA$^{-2}$]}$ \cite{zhu2005modern}. The magnetic field dependence shown in Fig. \ref{fig2}(c) allows us to extract the in-plane magnetization $M_y$ of the sample which can be used as a known parameter for micromagnetic simulations of the sub-nanometer-thick CoFeB layers. The shape of the $M_y(\mu_0 H)$ curve is in an excellent agreement with the scaled in-plane Alternating Gradient Field Magnetometry (AGFM) data also shown in Fig.~\ref{fig2}(c). The saturation magnetization of the CoFeB layers is found to be $M_\textrm{s}= 920\pm73$\,kA/m, which is smaller than that found in a previous study ($M_\textrm{s}\approx1200$\,kA/m) \cite{conca2013low,srivastava2023resonant}. Since the exact value of saturation magnetization depends on the precise choice of sputtering conditions, choice of the neighboring layers, quality of interfaces, processing protocols, etc. \cite{gowtham2016thickness}, a somewhat decreased $M_\textrm{s}$ in sub-nanometer-thick layers of CoFeB is not surprising \cite{gowtham2016thickness,lee2017temperature}.

\subsection{Evolution of the magnetic structure}

SANS was applied to investigate the evolution of the magnetic textures in the stack of MMLs as a function of applied magnetic field and temperature. At zero field and room temperatures, a ring-like SANS pattern, corresponding to orientationally-disordered stripe domains with an average period $l=118.3\pm0.5$\,nm was observed (data not shown). The MML studied here has been designed to have close to zero effective magnetic anisotropy ($<50$\,kJ/m$^3$ $<\mu_0 M_\mathrm{s}^2/2=0.53$\,MJ/m$^3$) at room temperature, this being responsible for larger domain walls (parametrized \cite{fallon2019quantitative} by $\pi \delta = \pi \sqrt{A/K}\approx40$\,nm) and shorter stripe periods compared to samples with larger effective anisotropy \cite{sampaio2013nucleation,moreau2016additive,legrand2018hybrid,burgos2023probing}, and hence a configuration that can stabilize a spin spiral In this case, an in-plane magnetic field saturation of the MML results, at remanence, in a stripe domain configuration, with the stripes extending along the applied field direction. In the present experimental geometry (Fig. \ref{fig1}(b)) the in-plane training field was applied after rotating the sample stick 90$^{\circ}$ around the $y$-axis. The expectation for a propagation vector of the Bloch-type spiral type is that it aligns with the direction of the external magnetic field after in-plane field-training \cite{bak1980theory,bauer2017symmetry}, while that of N\'eel-type cycloids aligns perpendicularly to the in-plane field \cite{bordacs2018magnetic,kurumaji2021direct}.

\begin{figure*}[ht]
\includegraphics[width=1\linewidth]{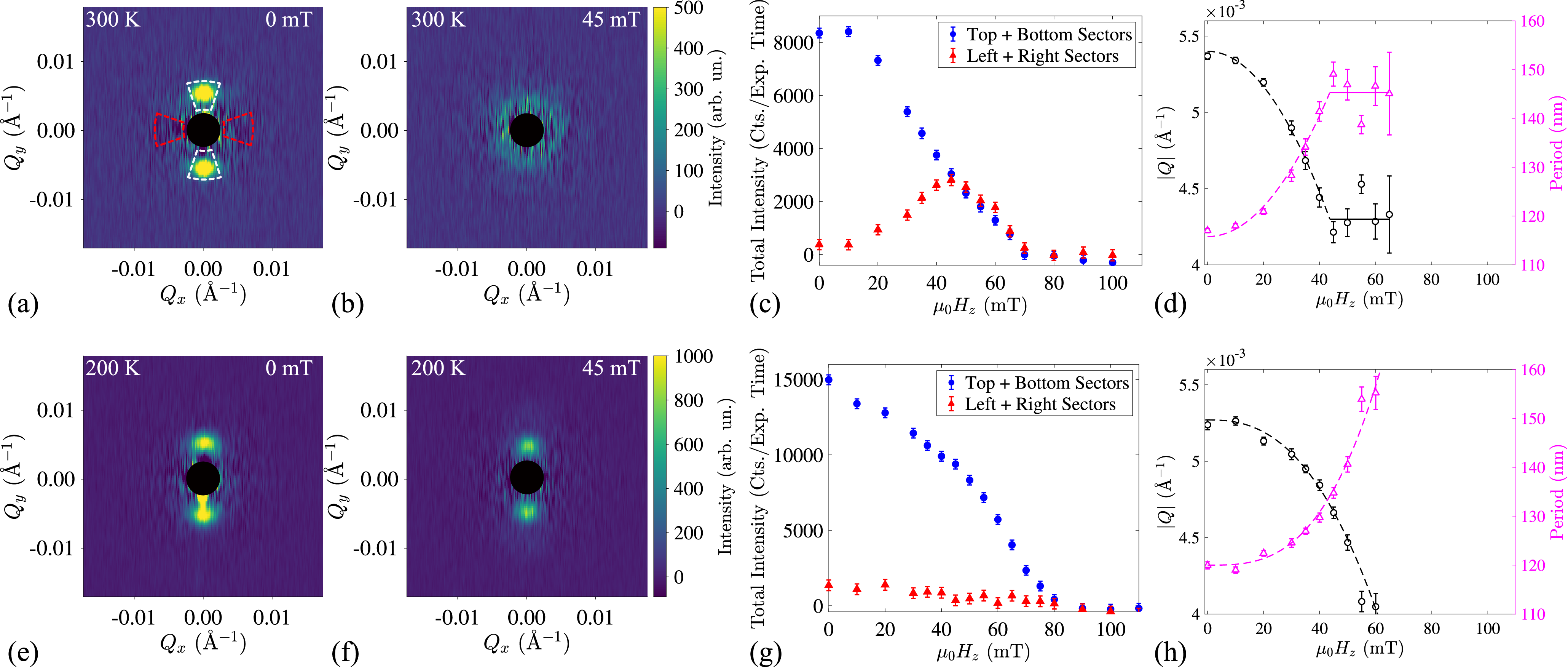}
        \caption{SANS pattern and intensity as a function of temperature and field. (a,b) SANS patterns measured at room temperature and in (a) zero applied magnetic field and (b) in 45\,mT. The low-$Q$ region containing the imperfect background subtraction around the beamstop is masked by the black disc. (c) Magnetic field dependence of the integrated intensity at $T=300$\,K. Blue and red dots correspond to the intensity integrated in 30$^\circ$-wide white and red sectors shown in the panel (a), oriented along $Q_y$ and $Q_x$, respectively. (d) Magnetic field dependence of the spiral or the skyrmion lattice $Q$ amplitude and the corresponding real-space periodicity at room temperature. Solid lines are guides to the eye. Bottom panels (e-h) correspond to analogous SANS dataset measured at a different temperature ($T=200$\,K). Here, no signature of the spiral to skyrmion phase transition is observed.}
        \label{fig3}
\end{figure*}

At remanence, the wavevector of the stripes is always along $\mathbf{Q_y}$. This was confirmed experimentally using SANS after rotating the sample back so that the neutron beam was incident along the out-of-plane direction of the MML. Fig. \ref{fig3}(a) shows the resulting SANS pattern with two strong spots along the $\pm\mathbf{Q_y}$ directions, as expected for the oriented stripe domains. Such a field-training procedure was implemented prior to every out-of-plane field ($\mu_0 H_z$) scan.

At room temperature (300~K), as the out-of-plane field was increased, the stripe domains were observed to transform gradually into a skyrmion phase. This manifests as a ring-like intensity distribution arising from orientationally disordered yet sufficiently-dense skyrmion arrangement [Fig. \ref{fig3}(b)]. The absence of the long-range orientational order is not surprising in a heavy metal / ferromagnet MMLs \cite{zazvorka2020skyrmion}, with the notable exception of Fe/Gd multilayers that display a well-defined hexagonal lattice arrangement when nucleated from the ordered stripe phase \cite{desautels2019realization}. Nevertheless, the magnetic phase transition from the stripe domain phase into the skyrmion phase is inferred clearly in the magnetic field dependence of SANS intensities in the sectors integrated along $Q_y$ and $Q_x$ projections of the momentum transfer vector, respectively shown in Fig. \ref{fig3}(a) as the white and red sector pairs. The transformation of the stripe domain texture into skyrmion state occurs gradually from zero field, and is complete at $\sim 45$\,mT. According to Fig. \ref{fig3}(c), at $\sim 45$\,mT, the intensities in the various sector boxes become equal, consistent with the observation of an isotropic ring. Beyond $\sim 45$\,mT the intensities in both types of sectors remain balanced, and decrease at the same rate until the total SANS signal vanishes completely at the saturation field of $80$\,mT, where the sample is magnetized homogeneously. The transition from stripe domains into skyrmions also shows up clearly in the field dependence of the magnetic propagation vector $|Q|$ (or spiral period $l=2\pi/Q$) which shows a continuous decrease (increase) with increasing field, before the continuous phase transition $\sim 45$\,mT whereafter $|Q|$ remains constant within the skyrmion phase (Fig. \ref{fig3}(d)). This $H$-dependent behaviour of $Q$ is typical for spiral magnets as a result of the competition between exchange interaction, DMI, anisotropies and external magnetic field  \cite{mcgrouther2016internal,burn2020field,srivastava2023resonant}.

\begin{figure*}
\includegraphics[width=1\linewidth]{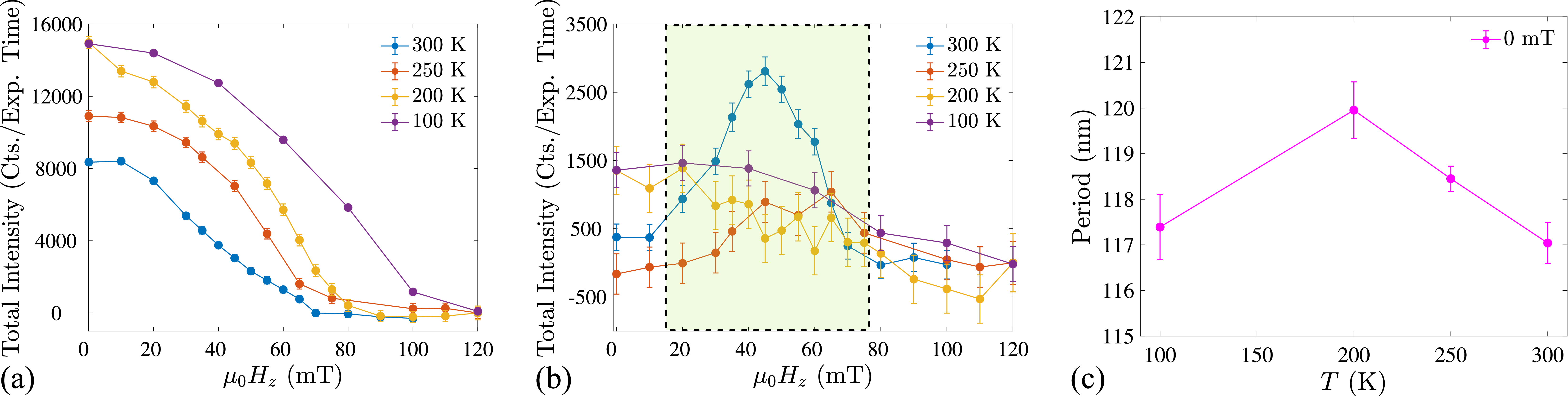}
        \caption{Magnetic field- and temperature-dependence of the total SANS intensity in (a) the top and bottom sectors, corresponding to the stripe domain peaks, and (b) the left and right sectors, corresponding to the skyrmion signal. The magnetic field region where skyrmions are located at 250\,K and 300\,K is shaded. (c) Temperature-dependence of the zero-field period of the spiral.}
        \label{fig4}
\end{figure*}

As the temperature is lowered to $T=200$\,K the SANS signature of the skyrmion phase becomes completely suppressed. Figs.~\ref{fig3}(e)-(h) show SANS patterns [Fig.~\ref{fig3}(e),(f)] measured at $\mu_0 H_z = 0$ and 45\,mT, and the magnetic field dependencies of the integrated intensity [Fig.~\ref{fig3}(g)] and $|Q|$ [Fig.~\ref{fig3}(h)] at $T=200$\,K. Here, the SANS peaks from the stripe domains formed after field-training are gradually suppressed by the applied field until the sample is saturated at $\mu_0 H_z\approx 80$\,mT. No isotropic intensity ring denoting skyrmion formation is observed over the whole magnetic field range [Fig.~\ref{fig3}(f),(g)]. While the magnitude of the wavevector (and the spiral periodicity) at zero field remain similar to that at 300\,K, Fig.~\ref{fig3}(h) shows $Q(\mu_0 H_z)$ to decrease monotonously without the transition to a plateau-regime characteristic of the skyrmion phase seen in the 300~K data. Figs.~\ref{fig4}(a) and (b) summarize the magnetic field- and temperature-dependencies of the total SANS intensity in the $Q$-space regions corresponding to the spiral [Figs.~\ref{fig4}(a)] and skyrmion [Figs.~\ref{fig4}(b)] phases at temperatures between 100\,K and 300\,K. The data suggest that skyrmions are absent at 100\,K and 200\,K, their weak signature exists only at 250\,K, while they are most pronounced at 300\,K.

In contrast to bulk skyrmion-hosting materials \cite{tokura2020magnetic}, the magnetic field vs. temperature phase diagrams of multilayered systems are scarcely reported. To our knowledge, the observed destabilisation of the skyrmion phase in the MML upon cooling below 250~K has hitherto not been reported experimentally. Indeed, according to the SANS data, Fig. \ref{fig4}(c) shows that the magnitude of the spiral $Q$ vector is almost temperature-independent, which indicates the weak thermal variation of the principal magnetic interactions in the system, namely, exchange, DMI, and dipolar interactions. The role of thermal fluctuations, which is important for the skyrmion stability in bulk noncentrosymmetric magnets \cite{muhlbauer2009skyrmion}, should be negligible in the present case, since the Curie temperature of $\sim 1$\,nm-thick CoFeB is well above room temperature ($T_\textrm{C} \approx 750$\,K \cite{lee2017temperature}).

Instead, it has been shown that the effective out-of-plane anisotropy energy in CoFeB increases significantly with decreasing temperature \cite{lee2017temperature}, leading us to suggest that an increase of the anisotropy below 250\,K is responsible for the suppression of the dense skyrmion state. In theoretical work, it is found that an increased effective anisotropy destabilizes the skyrmion lattice phase, and the ground-state spiral texture continuously transforms into the field-polarised phase containing isolated skyrmions within saturated state \cite{kiselev2011chiral,leonov2016properties}. Furthermore, the concomitant decrease in skyrmion density has been confirmed in a number of experimental studies on MMLs with tailored anisotropies \cite{soumyanarayanan2017tunable,he2018evolution,zhang2018creation}. Sparse magnetic objects such as an isolated skyrmions would only contribute the form factor into the SANS cross-section \cite{muhlbauer2019magnetic}, which is not detectable in the present experimental technique due to the extremely small sample volume. Therefore, further investigations of the low temperature vs. magnetic field phase diagram of the MMLs will benefit from real-space probes, such as magnetic force microscopy, x-ray microscopy, or Lorentz transmission electron microscopy, as well as detailed measurements of the temperature dependence of the magnetic anisotropy.

\subsection{Pure N\'eel-type domain walls}

To clarify the winding-type of the magnetic textures in detail, i.e. to both detect and discriminate between N\'eel-type and Bloch-type windings respectively stabilized by DMI and dipolar interactions, a SANS experiment with polarization analysis has been performed. For this experiment, we applied the same field-training procedure to the MML stack as for unpolarised SANS measurements, before applying a small out-of-plane guide field $\mu_0 H_z = 20$\,mT for the neutrons in the stripe domain state (see Fig.\ref{fig1}(b)). In a second step, the field was increased to $\mu_0 H_z = 45$\,mT to stabilize the skyrmion lattice. According to the selection rules for longitudinal polarisation analysis, neutrons with polarization $\pm \mathbf{P}_0$ along the incident beam with wavevector $\mathbf{k}_i$ (perpendicular to the MML plane) scatter from the N\'eel-type domain walls without changing their polarization \cite{maleev1963scattering,blume1963polarization,moon1969polarization}. Therefore, the intensity of the corresponding magnetic SANS signal will be observed only in the non-spin-flip (NSF) channel. On the other hand, Bloch-type domain walls cause the precession of the neutron spin, such that equal intensities will be observed in both the NSF and spin-flip (SF) channels. The general expressions for NSF and SF cross-sections are
\begin{eqnarray} \label{eq1}
    I_\textrm{NSF} = I^{++}+I^{--} \propto |2 m_z ^\perp (\m Q)|^2, \nonumber \\
    I_\textrm{SF}=I^{+-}+I^{-+} \propto | m_x ^\perp (\m Q) + \textrm{i} m_y ^\perp (\m Q) |^2  \nonumber \\ 
    + | m_x ^\perp (\m Q) - \textrm{i} m_y ^\perp (\m Q) |^2,
\end{eqnarray}
where $ m_x ^\perp (\m Q)$,  $ m_y ^\perp (\m Q)$, and  $ m_z ^\perp (\m Q)$ are the Fourier transform components of the local magnetization vector $\m m$ normal to $\m Q$, and  $x$, $y$, $z$ are Cartesian coordinates with the $z$ axis parallel to the wavevector of the incoming beam $\mathbf{k}_i$ and $\mathbf{P}_0$, and $x,y$ being in the sample plane. Previously, the polarised SANS technique has been used successfully for identifying the character of incommensurate modulations in bulk polar magnets GaV$_4$S$_8$ \cite{white2018direct} and VOSe$_2$O$_5$ \cite{kurumaji2021direct} with N\'eel-type skyrmions and cycloids, and centrosymmetric RKKY magnets EuAl$_4$ \cite{takagi2022square} and Gd$_2$PdSi$_3$\cite{ju2023polarized} with Bloch-type skyrmions and proper-screw helices. Note, that the present experimental geometry does not provide access to the absolute winding-handedness of the spin textures, which, in principle, can be determined if $\pm \mathbf{P}_0$ is oriented collinear to $\mathbf{Q}$ \cite{grigoriev2006magnetic,grigoriev2009crystal}.

\begin{figure*}
\includegraphics[width=1\linewidth]{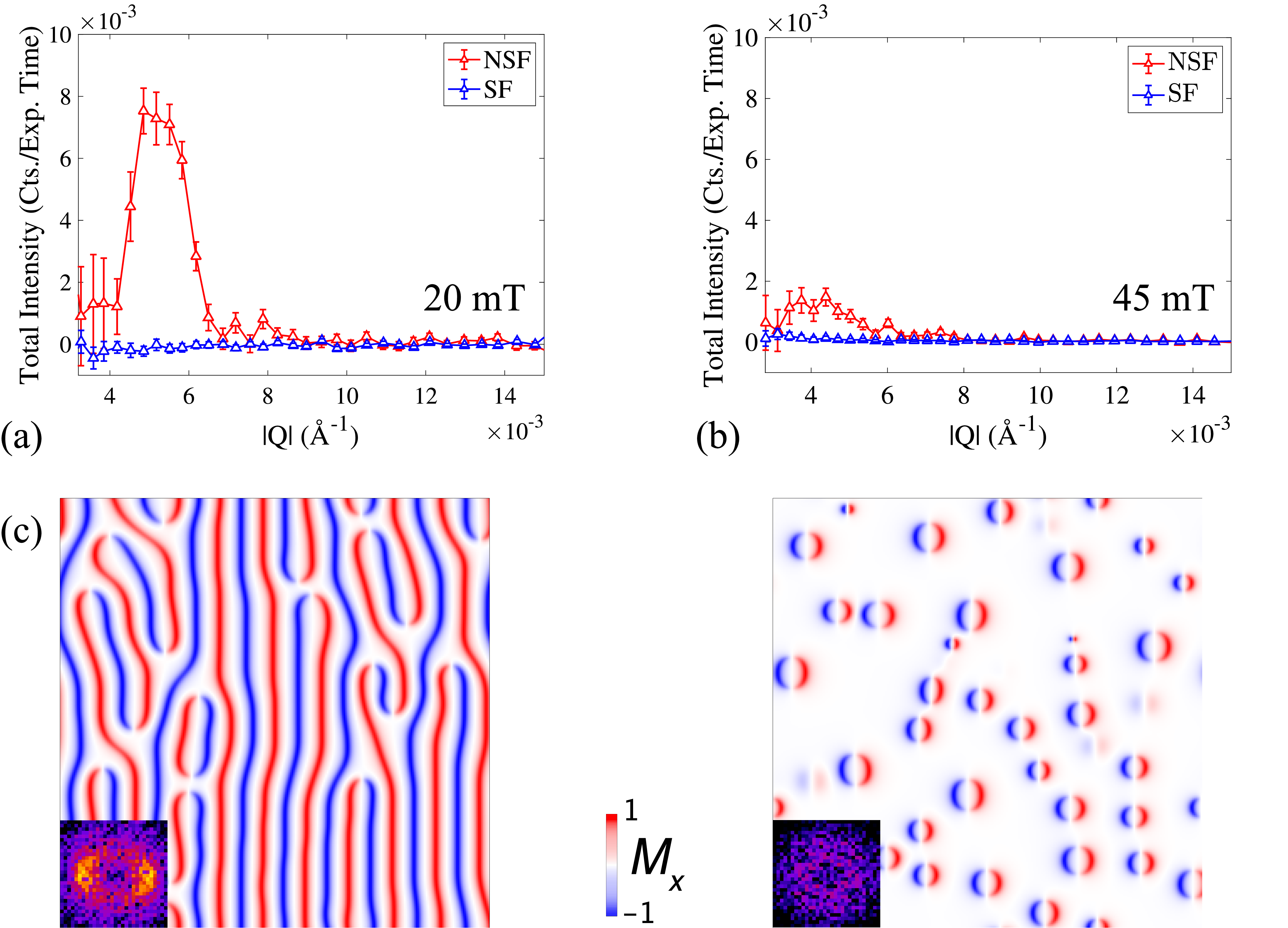}
        \caption{Integrated intensity profiles of non-spin-flipped (NSF) and spin-flipped (SF) SANS patterns measured at room temperature and (a) guide magnetic field of 20\,mT in the cycloidal phase and (b) 45\,mT in the skyrmion phase. Intensities were integrated over cycloidal Bragg peaks, and over complete SANS ring in the cycloidal and skyrmion phases, respectively.} (c) Aligned cycloidal and disordered N\'eel-type skyrmion textures obtained from the micromagnetic modelling. Cycloids were aligned using an in-plane (vertical) magnetic field training. In-plane magnetization component $M_x$ is shown. Fast Fourier transform (FFT) of the patterns are shown as insets. The simulation size is $256\times256\times19$ cells with the voxel size of $4\times4\times1.9$\,nm$^3$.
        \label{fig5}
\end{figure*}

The result of the present SANS experiment with polarization analysis is summarized in Fig. \ref{fig5}. It is clear that in both the spiral phase at an applied guide magnetic field of 20\,mT [Fig. \ref{fig5}(a)] and in the skyrmion phase at the field of 45\,mT [Fig. \ref{fig5}(b)], the observed SANS intensities are concentrated only in the NSF channel, while the SF intensities are zero within the errorbars. The ratio between SF and NSF integrated peak intensities are of an order of $10^{-5}$. This allows us to conclude that the magnetic textures observed by neutrons throughout the entire MML are uniquely described by N\'eel-type domain walls, and hence can be ascribed as cycloidal modulations and N\'eel-type skyrmions. The absence of any detectable SF intensity rules out the existence of a helical-winding admixture, as may be expected for mixed cycloidal-helical windings of the spin textures. These conclusions agree well with theoretical proposals that Bloch-type and hybrid domain walls are confined to the inner layers of relatively thick MMLs, while in thinner films purely N\'eel-type cycloids and skyrmions are stabilized the by the interfacial DMI.

Our microscopic observations of purely N\'eel-type domain walls are furthermore consistent with micromagnetic calculations performed using MuMax3 \cite{vansteenkiste2014design,exl2019preconditioned}. Indeed, looking at the three dimensional stripe magnetization textures for which our observed average period of 118\,nm is an energy minimum, we find only homochiral domain walls in the range of parameters corresponding to our experiments. The only hybrid walls are found for parameter sets that look unlikely, namely for $M_\mathrm{s}=1000$\,kA/m (upper bound) and the uniaxial out-of-plane anisotropy in its lower bound $K_\mathrm{u}=800$\,kJ/m$^3$ when the symmetric exchange $A_\mathrm{ex}$ is $<5$\,pJ/m. Moreover, even in this hybrid case, the walls in each individual CoFeB layers are still mostly N\'eel-type, with the top layer switching from counter-clockwise to clockwise N\'eel configuration, without substantial Bloch component. Illustrative spin textures obtained from the micromagnetic model at zero field after the in-plane field training, and in out-of-plane field of 65\,mT using $K_\mathrm{u}=660$\,kJ/m$^3$, $M_\mathrm{s}=920$\,kA/m, $A_\mathrm{ex}=4$\,pJ/m, and interfacial DMI constant $D=0.8$\,mJ/m$^2$ are shown in Fig. \ref{fig5}c. Both cycloidal and skyrmion textures were obtained using in-field relaxation of the initially random spin textures in the presence of an in-plane and out-of-plane magnetic fields, respectively. The in-plane projection $M_x$ of magnetization allows the unambiguous distinction between Néel and Bloch type domain walls. \cite{luo2023direct}. The obtained micromagnetic parameters reproduce well the aligned cycloidal textures with the period of 118\,nm, and the field-induced N\'eel skyrmion phase.

\section{Conclusions}

In summary, we have presented a microscopic study of the magnetic order throughout entire [Pt(1 nm)/(CoFeB(0.8 nm)/Ru(1.4 nm)]$_{10}$ multilayers using the polarised neutron reflectometry (PNR) and small-angle neutron scattering (SANS) techniques. By using SANS and PNR, we were able to probe the magnetic order in the samples with both bulk and layer-resolved sensitivities, respectively. The data reveal the existence of long-periodic magnetic stripe domains and orientationally disordered skyrmion phases in the multilayers. We find that the skyrmion phase in the multilayer is suppressed below $T\approx250$\,K, which is likely to be the consequence of the temperature dependence of the effective anisotropy. Using polarised SANS we unambiguously identified a pure N\'eel-type character of the magnetic modulation in [Pt(1 nm)/(CoFeB(0.8 nm)/Ru(1.4 nm)]$_{10}$ with no detectable Bloch-type admixture. These proof-of-principle experiments pave the way towards investigating more complex phenomena in MMLs by mesoscale SANS probes, such as detecting the existence and layer-specific locations of hybrid Bloch-N\'eel windings of the magnetic textures. Besides being a powerful probe of the bulk magnetic structure, experiments performed in the small-angle scattering geometry allow the extraction of microscopic parameters in DMI-driven helicoidal \cite{grigoriev2015spin} and cycloidal \cite{utesov2022small} systems, such as spin wave stiffness and damping from the quasi-elastic spin wave scattering \cite{ukleev2022spin}. Overall, our study provides a pathway towards a deeper microscopic understanding of magnetic order in skyrmion-hosting MMLs, which is crucial for the development of future spintronic devices based on skyrmions. 

\section*{Acknowledgement}
SANS and PNR experiments were carried out respectively using D33 and SuperADAM instruments a as a part of proposals 5-54-305 \cite{illdata1}, 5-54-333 \cite{illdata2}, 5-54-379 \cite{illdata3}. We thank O. Aguettaz and M. Bonnaud for the technical assistance during the beamtimes, and B.P. Toperverg for fruitful discussions.

\section*{Disclosure statement}

Authors declare no conflict of interest.

\section*{Funding}

V.U. and J.S.W. acknowledge funding from the Swiss National Science Foundation (SNSF) Projects 200021\_188707 and Sinergia CRSII5\_171003 NanoSkyrmionics. V.U., C.L., R.A., F.R. acknowledge financial support by the German Federal Ministry for Education and Research (BMBF project No. 05K19W061) and financial support by the German Research Foundation via Project No. SPP2137/RA 3570. F.A., V.C. and N.R. acknowledge France 2030 government grant managed by the French National Research Agency (ANR-22-EXSP-0002 PEPR SPIN CHIREX), the French National Research Agency (ANR) with ANR TOPO3D (ANR-22-CE92-0082) and the DARPA TEE program (Grant No. MIPRHR-0011831554). F.A. acknowledges Comunidad de Madrid  Atracción de Talento program (Grant No. 2022-T1/IND-23901).

\section{Appendix A: Transition from the spiral to the skyrmion phase}

Detailed SANS data on the transition from the aligned spiral to the skyrmion state is shown in Figs. \ref{fig7}a--d. The visible ring-like scattering originated from disordered skyrmions emerges at ca. 35 mT and co-exists with cycloidal Bragg peaks. Upon further field increment only the ring remains. Radial profiles of the SANS intensity in cycloidal and skyrmion phases are shown in Fig. \ref{fig7}e. Distinctly from crystalline skyrmion-hosting materials and dipolar skyrmions in Fe/Gd multilayers, the azimuthal distribution of SANS intensity in the skyrmion phase does not have a six- or twelve-fold symmetry (Fig. \ref{fig7}f).

\section{Appendix B: Anomalous Hall effect}
The electric transport measurements were performed using the Alice II chamber at the BESSY-II synchrotron \cite{abrudan2015alice,ukleev2022chiral}. The temperature dependence of the anomalous Hall effect (AHE) was studied by means of spinning-current anomalous Hall magnetometry \cite{kosub2015all} using a Tensormeter model RTM-1 (HZDR Innovation GmbH, Germany). Anomalous Hall resistance $R_{xy}$ of the multilayer was probed in a four-wire configuration using the zero-offset Hall scheme of the device. The sample temperature was controlled using the closed-cycle refrigerator (Stinger, ColdEdge Technologies, USA). The magnetic field generated by an electromagnet was applied perpendicular to the sample surface.

\begin{figure*}
\includegraphics[width=1\linewidth]{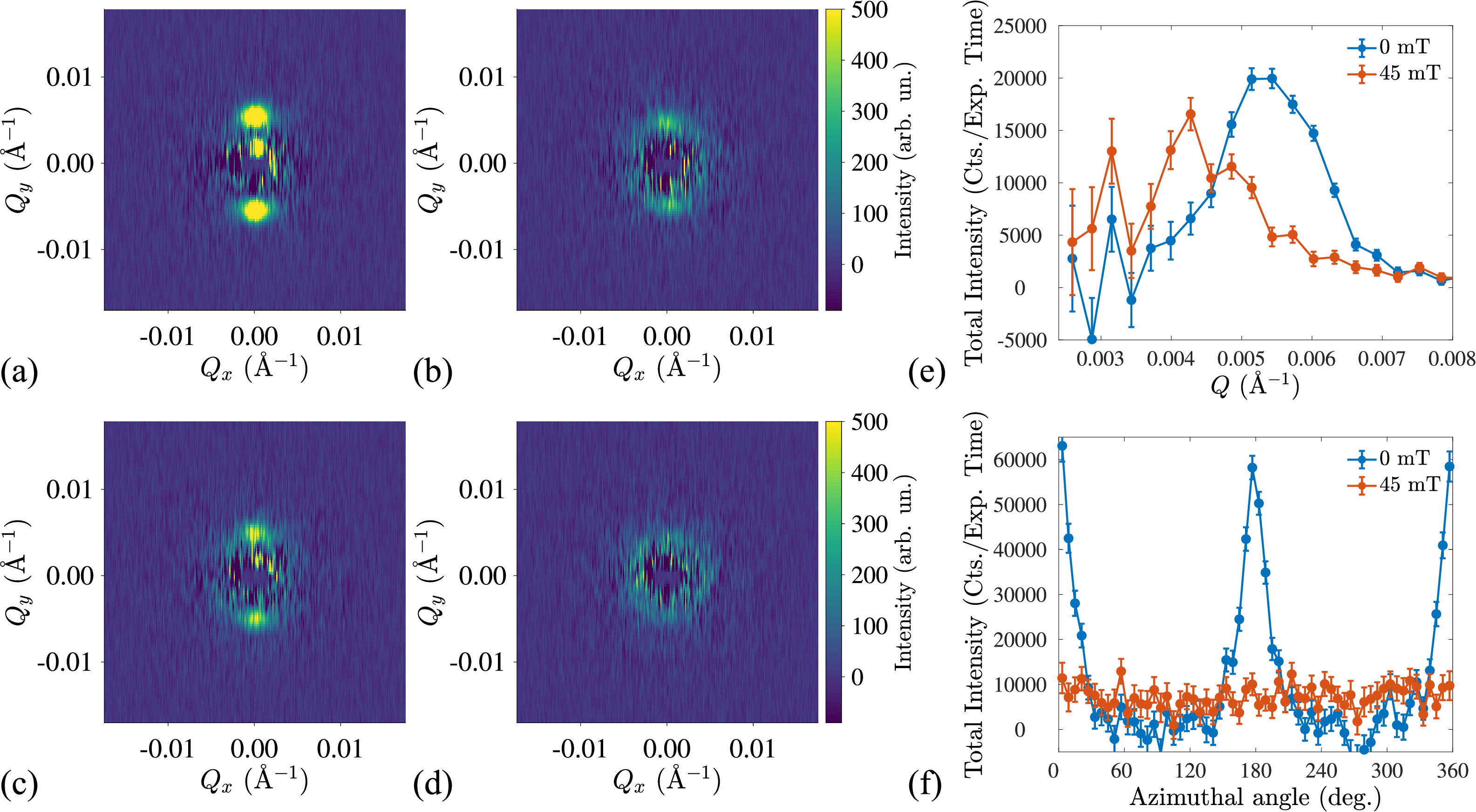}
        \caption{SANS patterns measured on the transition from cycloidal to skyrmion phases at (a) 0 mT, (b) 30 mT, (c) 35 mT, (d) 45 mT. (e,f) Radial and azimuthal profiles of the SANS intensity measured in cycloidal and skyrmion phases at 0 and 45 mT, respectively. }
        \label{fig6}
\end{figure*}

\begin{figure*}
\includegraphics[width=1\linewidth]{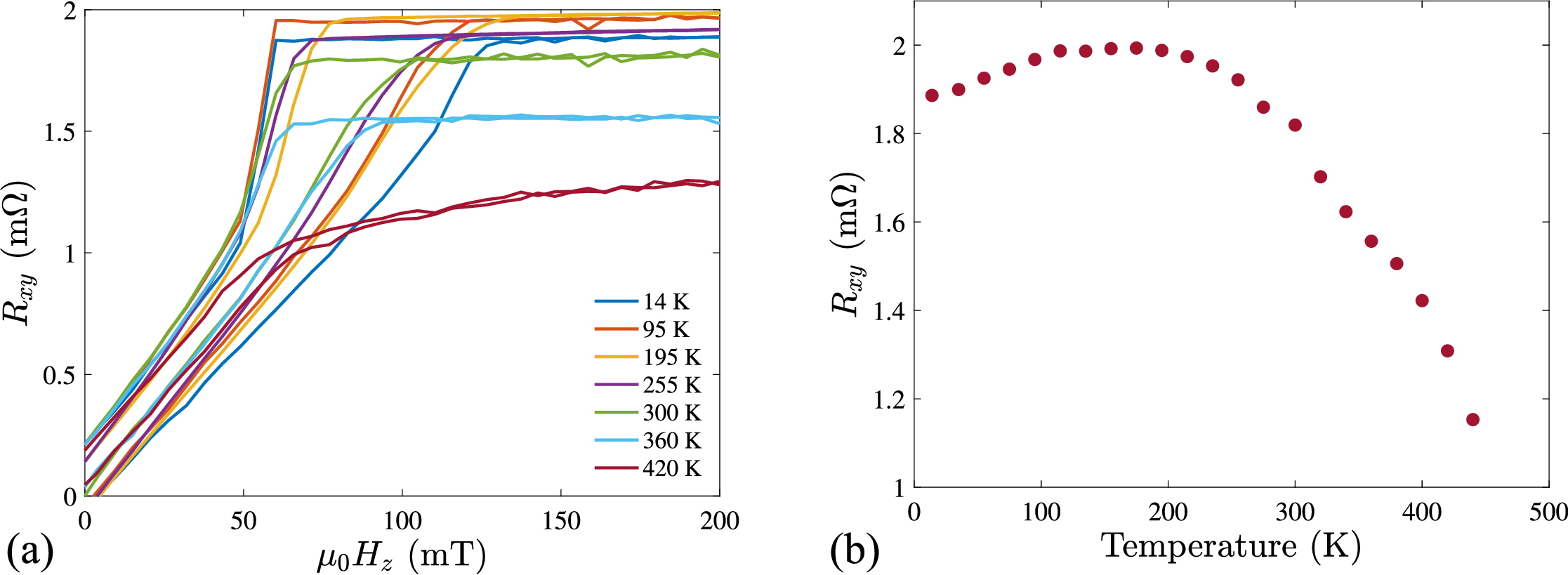}
        \caption{(a) Anomalous Hall resistance vs. out-of-plane magnetic field curves measured at different temperatures. (b) Temperature dependence of the Hall resistance of the saturated multilayer.}
        \label{fig7}
\end{figure*}

An additional facet of our investigation involved anomalous Hall resistance measurements. Selected anomalous Hall resistance curves are presented in Fig. \ref{fig6}a, revealing distinctive hysteresis loops that maintain a consistent shape across the temperature range of 14\,K to 360\,K. Notably, variations in the magnitude of the anomalous Hall effect (AHE) signal $R_{xy}$ are evident in the saturated state. Intriguingly, upon further warming up to 440\,K, a distinctive S-shaped hysteresis loop emerges in the magnetization behavior.

The temperature dependence of Hall resistance, as measured in the magnetically saturated state, is depicted in Fig. \ref{fig6}b. In the temperature range between 440\,K and approximately 200\,K, a consistent increase in the magnitude of $R_{xy}$ with decreasing temperature is observed.

This observation aligns with the absence of the skyrmion phase at temperatures below room temperature, as corroborated by the SANS experiments. However, a detailed discussion of the AHE including various scattering mechanisms in the MML \cite{dang2020anomalous} lies beyond the scope of the present paper.

\bibliographystyle{tfnlm}

\newpage

\end{document}